\documentclass[aip,pop,preprint,groupedaddress,preprintnumbers,showpacs,amsmath,amssymb]{revtex4-1}

\usepackage{graphicx}

\begin{document}

\title{Nonthermal electron and ion acceleration by magnetic reconnection in large laser-driven plasmas}

\author{S. R. Totorica} \email{totorica@princeton.edu} 
\affiliation{High Energy Density Science Division, SLAC National Accelerator Laboratory, Menlo Park, California 94025, USA}
\affiliation{Kavli Institute for Particle Astrophysics and Cosmology, Stanford University, Stanford, California 94025, USA}
\affiliation{Department of Astrophysical Sciences, Princeton University, Princeton, New Jersey 08544, USA}
\affiliation{International Research Collaboration Center, National Institute of Natural Sciences, Tokyo 105-0001, Japan}

\author{M. Hoshino} 
\affiliation{Department of Earth and Planetary Science, University of Tokyo, Tokyo 113-0033, Japan}

\author{T. Abel}
\affiliation{Kavli Institute for Particle Astrophysics and Cosmology, Stanford University, Stanford, California 94025, USA}
\affiliation{Department of Physics, Stanford University, Stanford, California 94305, USA}
\affiliation{SLAC National Accelerator Laboratory, Menlo Park, California 94025, USA}

\author{F. Fiuza} \email{fiuza@slac.stanford.edu} 
\affiliation{High Energy Density Science Division, SLAC National Accelerator Laboratory, Menlo Park, California 94025, USA}

\date{\today}

\begin{abstract} 
Magnetic reconnection is a fundamental plasma process that is thought
to play a key role in the production of nonthermal particles
associated with explosive phenomena in space physics and astrophysics.
Experiments at high-energy-density facilities are starting to probe
the microphysics of reconnection at high Lundquist numbers and large
system sizes. We have performed particle-in-cell (PIC) simulations to explore particle acceleration
for parameters relevant to laser-driven reconnection experiments. We
study particle acceleration in large system sizes that may be produced soon with the most energetic
laser drivers available, such as at the National Ignition Facility. In these
conditions, we show the possibility of reaching the multi-plasmoid
regime, where plasmoid acceleration becomes dominant.  Our results
show the transition from \textit{X} point to plasmoid-dominated acceleration associated with the merging and contraction of plasmoids that further extend the
maximum energy of the power-law tail of the particle distribution for
electrons. We also find for the first time a system-size-dependent
emergence of nonthermal ion acceleration in driven reconnection, where the
magnetization of ions at sufficiently large sizes allows them to be contained by the magnetic field and
energized by
direct \textit{X} point acceleration.  For feasible experimental conditions,
electrons and ions can attain energies of
$\epsilon_{max,e} / k_{B} T_{e} > 100$ and
$\epsilon_{max,i} / k_{B} T_{i} > 1000$.
Using PIC simulations with binary Monte Carlo
Coulomb collisions we study the impact of collisionality on plasmoid formation and particle acceleration. The implications of these results for
understanding the role reconnection plays in accelerating particles in space
physics and astrophysics are discussed.
\end{abstract}
\pacs{}
\maketitle

\section{INTRODUCTION}

Magnetic reconnection is a fundamental plasma process that converts
magnetic field energy into plasma flows, heating, and energetic particles
\cite{Zweibel2009}. It is believed to play an important role in the dynamics
of magnetized plasmas in a wide range of scenarios in space physics,
astrophysics, and laboratory nuclear fusion devices
\cite{Ji2011,Joglekar2014,Taylor1986}.  One aspect of reconnection that is
of particular interest is its ability to produce energetic particles
with nonthermal distributions, which are directly measured in association
with reconnection in the Earth's magnetosphere \cite{Moebius1983,Oieroset2002,Imada2005} and inferred to be present in
magnetized astrophysical objects from radiation spectra \cite{Hoshino2012}.
The complex multi-scale dynamics involved in reconnection and particle
acceleration in realistic systems make it a challenge to fully understand
the particle acceleration properties of reconnection, and this is currently
an active area of research.

Energetic laser facilities, including the Gekko XII laser at Osaka University, the OMEGA laser at the University of
Rochester, and the National Ignition Facility (NIF) at Lawrence Livermore National
Laboratory, can focus kJ to MJ
energies onto sub-millimeter spot sizes over nanosecond time scales and
create high-energy-density plasma states in the laboratory. These lasers can
ablate plasmas with keV temperatures and $\sim 1000$ km/s flows when focused
onto solid targets, potentially reaching regimes where collisional mean free
paths exceed the system size.  In sufficiently collisionless regimes, the
governing equations can be shown to be invariant to scaling transformations
\cite{Ryutov2000,Ryutov2012} that open the possibility of using these
experiments to gain insight into the reconnection dynamics occurring in
systems in space physics and astrophysics \cite{Remington2006}.

Over the past decade there has been
significant effort and progress in developing laser-driven plasma
experiments into a platform for studying the dynamics of reconnection in a
controlled laboratory setting
\cite{Nilson2006a,Li2007a,Nilson2008a,Willingale2010a,Zhong2010a,Dong2012a,Fiksel2014a,Rosenberg2015a,Rosenberg2015b,Pei2016,Kuramitsu2018,Morita2019}.
By focusing kilojoule per nanosecond lasers onto
solid targets, expanding plasma plumes are ablated, and reconnection can occur
between self-generated Biermann battery
\cite{Haines1997,Stamper1991,Gao2015,Matteucci2018} fields or
externally imposed magnetic fields \cite{Fiksel2014a}.
These experiments are in an interesting regime with strongly driven flows, large system sizes in terms of the microscopic scales, and the potential
to reach relativistic regimes \cite{Raymond2018}.
The plasma properties and electromagnetic fields in these experiments
can be measured using diagnostics including proton radiography, Thompson scattering,
and shadowgraphy.
Past experiments have measured features of reconnection including
changes in magnetic field topology \cite{Li2007a}, plasma heating \cite{Dong2012a}, and the formation of plasma jets \cite{Nilson2006a},
however it is challenging to access the full details of the
kinetic physics occurring in these systems with current diagnostics and several important aspects, including nonthermal particle acceleration, are not yet fully understood.

In previous work by several of us, we have used fully kinetic particle-in-cell (PIC) simulations to model
laser-driven plasma experiments and found the possibility of nonthermal
electron acceleration for the conditions and geometries of recently
performed experiments \cite{Totorica2016,Totorica2017}.  For these
conditions we found through particle tracking studies that the electrons
are energized primarily by direct acceleration at the \textit{X} points, with
plasmoid related acceleration contributing only minor additional energy
gain. 
Numerical studies by other groups have shown similar electron
acceleration \cite{Fox2017,Huang2017,Huang2018}.
Here we extend our study to larger system sizes in terms of
the microscopic ion skin depth $d_{i} = c / \omega_{pi}$, which could be produced
at megajoule class laser facilities such as NIF.
Recent laboratory astrophysics experiments on NIF focused on the study of particle acceleration in collisionless shocks have demonstrated the generation of large collisionless plasmas with $L/d_{i} > 500$
for the system size \cite{Fiuza2020}. NIF experiments on magnetic reconnection are also underway and recent measurements have
suggested fast reconnection in a thin and elongated current sheet
\cite{Fox2020}. 
As the characteristic length scale of plasmoid formation is on the order of
the ion skin depth $d_{i}$, system sizes of $L / d_{i} > 100$ are
expected to reach a regime where many plasmoids are present in the current
sheet, potentially merging to larger scales and introducing new particle
acceleration channels.  We present below simulation results for such
system sizes. We find in our simulations that multi-plasmoid dynamics
enhance electron
acceleration, and we observe for the first time a system-size-dependent emergence of nonthermal ion acceleration in strongly driven systems for sufficiently
large sizes.  In the context of
laser-driven reconnection experiments, this is the first time that
nonthermal ion acceleration has been observed. Experiments in this regime would be valuable to benchmark numerical simulations in well-controlled conditions and would further extend our
understanding of the role reconnection plays in accelerating particles in
space and astrophysics, which are often expected
to be in the multi-plasmoid regime \cite{Ji2011}.

The outline of this manuscript is as follows. 
The details of the numerical simulations are given in Section II.
We investigate the acceleration of electrons in Section III 
and the acceleration of ions in Section IV, extending our previous studies
to larger system sizes.  In Section V we study the influence of particle
collisionality using collisional simulations. In Section VI we discuss
these results and their implications for reconnection in space
physics and astrophysics,
and in Section VII we summarize our conclusions.

\section{DESCRIPTION OF SIMULATIONS}

\begin{figure}[ht]
\begin{center}
\includegraphics[width=0.9\textwidth]{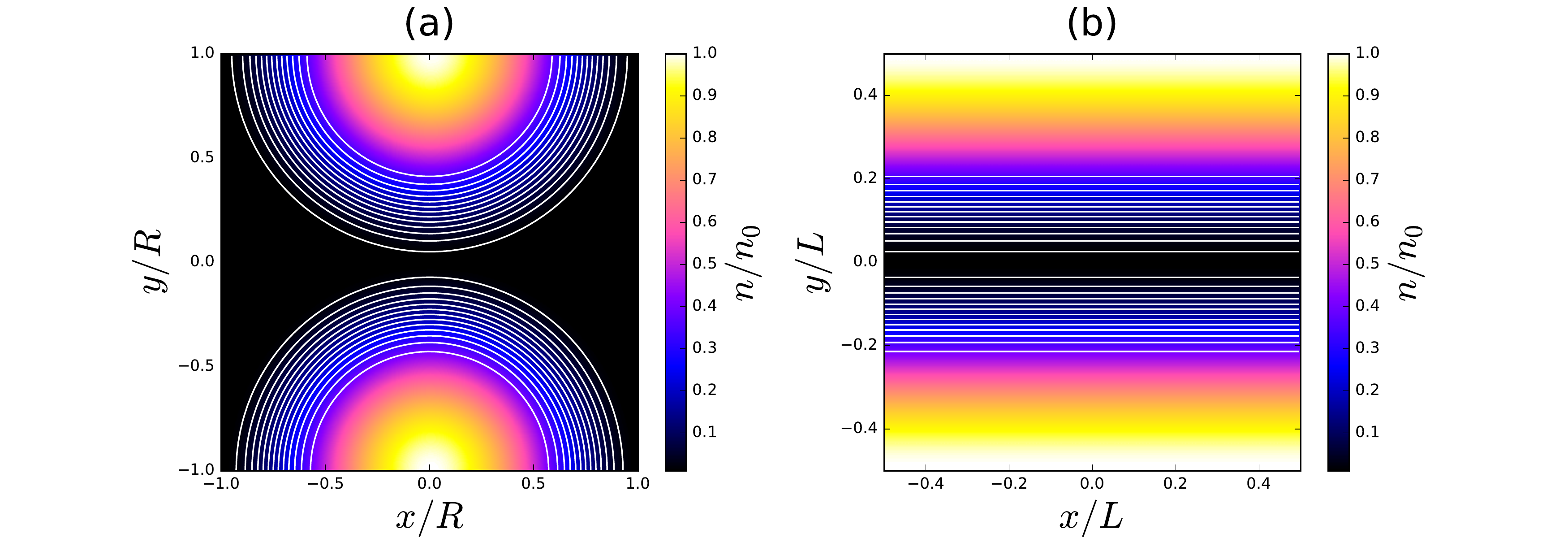}
\caption{\label{fig:setup} 
Density with overlaid magnetic fields lines (white) for the initial
conditions of the finite (a) and infinite (b) systems.
The finite system simulations have domains of $8 R \times 2 R$ with
bubble radius $R$ ranging from $R / d_{i} \simeq 26.5 - 106$.
The infinite systems have square domains of side length $L$ ranging
from $L / d_{i} \simeq 53 - 212$.
}
\end{center}
\end{figure}

To model the microphysics of laser-driven reconnection experiments we use
two-dimensional PIC simulations performed with the fully
relativistic, massively parallel, state-of-the-art PIC code
OSIRIS \cite{Hemker2015,Fonseca2002a,Fonseca2008,Fonseca2013}.
PIC simulations represent the plasma by a finite number of discrete
simulation particles that interact through self-consistent electromagnetic
fields \cite{Dawson1983a}, and can capture the full kinetic physics of
highly nonlinear plasma systems. We use both
standard collisionless PIC simulations as well as simulations with Coulomb
collisions taken into account through the use of a binary Monte Carlo
collision operator\cite{Takizuka1977}. The simulations are initialized at a
time partway through the experiment, when the two magnetized plasma bubbles
are expanding and about to interact.  These generic initial conditions can be connected to experimental geometries
including both self-generated and externally imposed magnetic fields
\cite{Ryutov2013a}, and have been used in previous PIC studies that have
given insight into laser-driven reconnection
\cite{Fox2011a,Fox2012a,Lu2013a,Lu2014,Totorica2016,Totorica2017}.
The centers of the plasma bubbles are
given by the vectors  ${\bf R}^{(1)} = (0, R, 0)$ and 
${\bf R}^{(2)} = (0, -R,
0)$, and the corresponding radial vectors from the bubble centers are ${\bf
  r}^{(i)} = {\bf r} - {\bf R}^{(i)}$, where $R$ is the radius of each
bubble when they begin to interact. The initial density profile is given by
$n_{b} + n^{(1)} + n^{(2)}$ where
$n^{(i)}(r^{(i)}) = (n_{0} - n_{b})\,\textup{cos}^{2}\left ( \frac{\pi r^{(i)}}{2 R} \right )$ if $r^{(i)} < R$, 0 otherwise.  Here,
$n_{b} = 0.01 n_{0}$ is the background plasma density.  The initial velocity profile
is given by ${\bf V}^{(1)} + {\bf V}^{(2)}$ where
${\bf V}^{(i)}(r^{(i)}) = V_{0}\,\textup{sin}\left ( \frac{\pi r^{(i)}}{R} \right ) {\bf r}^{(i)}$ if $r^{(i)} < R$, 0 otherwise. 
The initial magnetic field corresponds to the sum of two oppositely
aligned ribbons of finite flux, ${\bf B}^{(1)} + {\bf B}^{(2)}$ where $ {\bf
  B}^{(i)}(r^{(i)}) = B_{0} \, \textup{sin} \left (
  \frac{\pi (R - r^{(i)})}{2 L_{B}} \right ) {\bf
  \hat{\phi}}^{(i)}$ if $R - 2 L_{B} \leq r^{(i)} \leq R$, 0
otherwise.  Here $L_{B} = R / 4$ is the initial half-width of the
magnetic field ribbon. The initial electric field is 
${\bf E} = - {\bf V} \times {\bf B} / c$, consistent with the initial motion
of the magnetized plasma.  An initial out-of-plane current $J_{z}$ is
included that is consistent with
$\nabla \times {\bf B} = \frac{4 \pi}{c} {\bf J}$ and distributed to the
electrons and ions by the inverse of their mass ratio.  
In this study we use both systems that are uniform
and periodic along the direction of the reconnection outflows, as well as
finite systems with circular plasma bubbles, extending our previous work
to larger systems sizes in terms of the ion skin depth (Figure \ref{fig:setup}).  The periodic
systems allow us to study very large system sizes, and use the same initial
conditions as described above but with the radial vectors changed to
${\bf r}^{(1)} = (0, \; y - R,0)$ and ${\bf r}^{(2)} = (0, \; y + R,0)$.
These simulations have square domains with the length of each side ranging
from $L / d_{i} \simeq 53 - 212$.  The finite sized circular bubble systems
allow the inclusion of the effects of particle escape and for these we use
bubble radii ranging from $R / d_{i} \simeq 26.5 - 106$.
The anti-symmetry of these systems allows the use of periodic boundaries
along
the inflow ($y$) direction.  Periodic boundaries are also used along
the $x$ axis, and for the finite sized systems the domain is set to be 8
times the bubble radius along this axis to prevent recirculation of the
reconnection outflows.

We directly match many of the plasma parameters to their values estimated
from experimental measurements
\cite{Fox2012a}. For the Alfv\'{e}nic and sonic Mach numbers we model a
range of $M_{A} = V_{0} / V_{A} = 4 - 16$ and $M_{S} = V_{0} / C_{S} = 2$,
where $V_{A} = B_{0} / \sqrt{4 \pi n_{0} m_{i}}$ is the Alfv\'{e}n speed
and $C_{S} = \sqrt{Z T_{e} / m_{i}}$ is the sound speed. 
For the electron plasma beta this corresponds to $\beta_{e} = n_{0} T_{e} / (B_{0}^{2} / 8 \pi) = 2 (M_{A} / M_{S})^{2} = 8 - 32$. The initial
temperatures $T_{e}$ and $T_{i}$ are taken to be equal and uniform
throughout the plasma.  To make the simulations computationally feasible we
do not match the
experimental values for the parameters $C_{S} / c$ and $m_{i} / m_{e} Z$
(where $c$ is the speed of light, $m_{i}$ and $m_{e}$ the ion and electron masses, and
$Z$ the ion charge). For these parameters we use
use $V_{0} / c = 0.1$ and $m_{i} / m_{e} Z = 128$, with $C_{S}$ chosen
to correctly match $M_{S}$.
This is a commonly used approximation that
effectively reduces the speed of light
and increases the electron mass, but can still allow for an adequate
separation of the relevant temporal and spatial scales to accurately model
the physical processes \cite{Drake2006b,Fox2012a,Huang2017,Li2019}. 
This will primarily impact the electron-scale physics and electromagnetic
waves which are not the focus of this study, and we have performed
convergence tests to ensure this choice of numerical parameters is not obscuring
the physical processes \cite{Totorica2017}.
For the numerical parameters of the simulations,
the discrete timestep is $\Delta t = 0.35 \; \omega_{pe}^{-1}$, and the
resolution of the spatial grid is
$\Delta x = 0.5 c / \omega_{pe} \simeq 0.04 c / \omega_{pi}$. Cubic
interpolation is used to weigh between the particle and field
quantities, and 64 particles per cell per species are used in the initial
conditions.

\section{ELECTRON ACCELERATION}

\begin{figure}[ht]
\begin{center}
\includegraphics[width=0.9\textwidth]{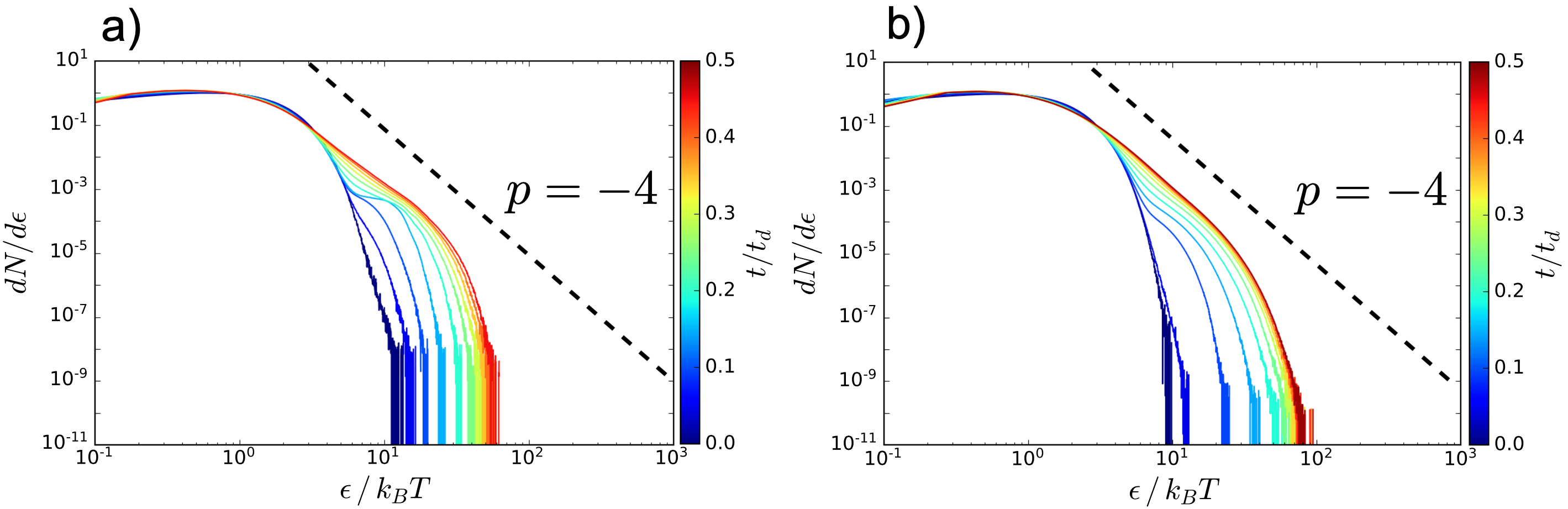}
\caption{\label{fig:electron_spectra} 
Temporal evolution of the electron energy spectra for two different system
sizes: (a) $L / d_{i} = 53$ and (b) $L / d_{i} = 212$.  The colored lines show the
evolution of the spectra from $t / t_{d} = 0$ to $t / t_{d} = 0.5$ at
equally spaced time intervals.   A power-law spectrum with index $p=-4$
is plotted for reference.
}
\end{center}
\end{figure}

We first discuss the details of electron acceleration, extending our
previous work \cite{Totorica2016,Totorica2017} to larger system sizes.
Figures \ref{fig:electron_spectra} (a) and (b) show the temporal evolution
of the electron spectra for systems with lengths along the reconnection
outflow direction ($x$) of $L / d_{i} = 53$ and $L / d_{i} = 212$,
respectively.   The plots show the temporal evolution
of the spectra from $t / t_{d} = 0$ to $t / t_{d} = 0.5$, where
$t_{d} = R / V_{0}$ is the relevant timescale for the interaction and the
experimental measurements. Both systems are periodic in the
$x$-direction and form plasmoids in the current sheet.  The system shown in
(b) is scaled four times larger along each dimension compared to the system shown in (a), and thus also
advects
four times the amount of magnetic flux into the current sheet.  This allows
reconnection to proceed for longer, leading to plasmoids growing and merging
which has important consequences for the particle acceleration.
The spectra in both cases are very similar, with a power-law index of
approximately $p=-4$.  Both the value of this index and the relative
insensitivity of the spectrum to system size are consistent with recent
measurements in the context of reconnection in Earth's magnetotail \cite{Imada2015a}.  The energy contained above 5 (10) times the thermal
energy is approximately 10 (2) percent of the total energy of the electrons
for the system with $L / d_{i} = 212$.

\begin{figure}[ht]
\begin{center}
\includegraphics[width=0.75\textwidth]{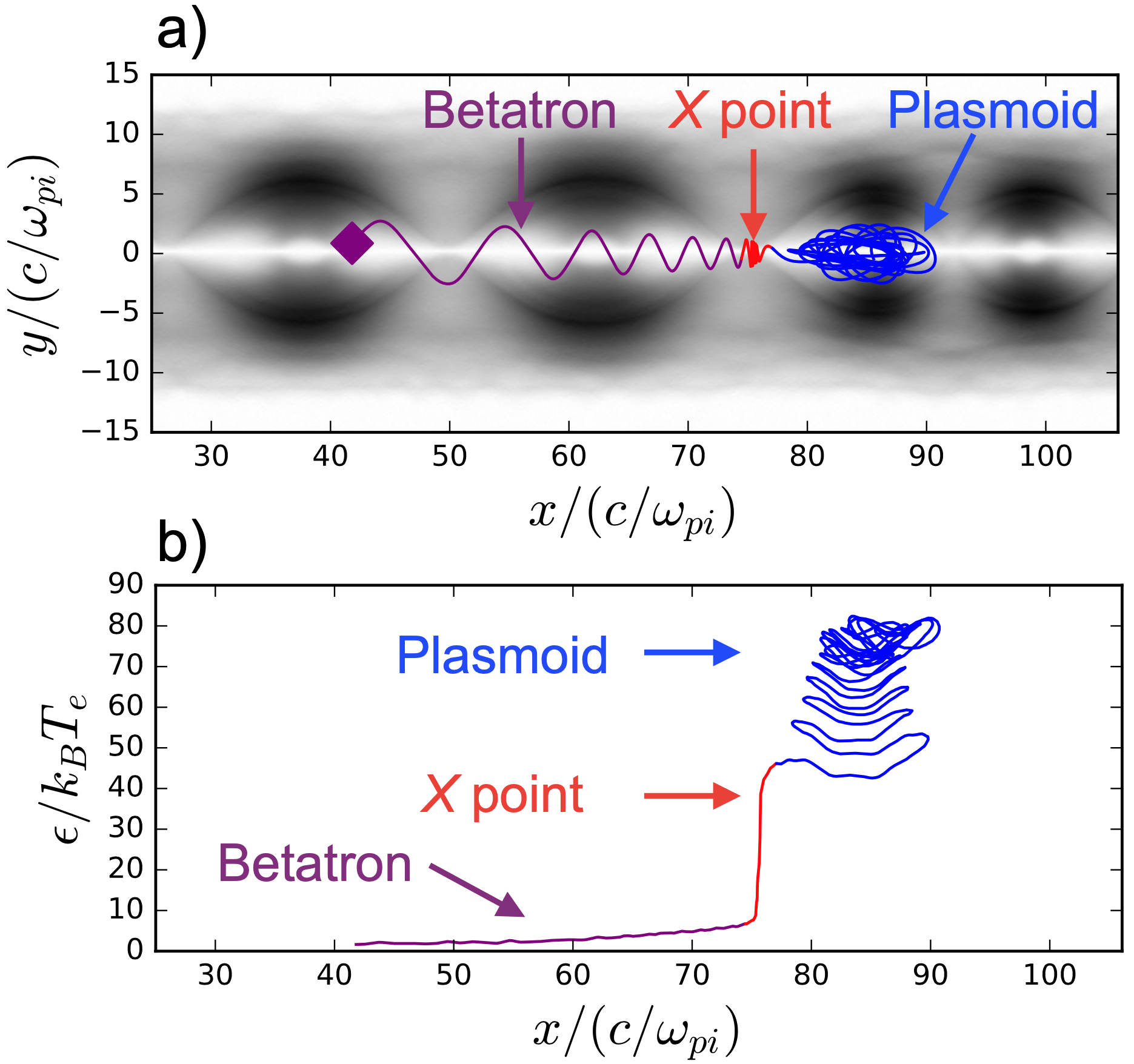}
\caption{\label{fig:three_phases} 
(a) Trajectory of a representative energetic electron from the
$L / d_{i} = 212$ simulation plotted over the 
magnitude of the in-plane magnetic field, with its initial position indicated
by the diamond. (b) Energy of the electron as a function of its position
along the $x$ axis.  The trajectories show the three different phases to the 
acceleration, distinguished by different colors.
}
\end{center}
\end{figure}

Figure \ref{fig:three_phases} shows an example trajectory of an energetic 
electron from the simulation with $L / d_{i} = 212$, highlighting the
importance
of energy gain inside the plasmoid. Figure \ref{fig:three_phases} (a) shows
the trajectory of the electron in space, plotted over the magnitude of the
magnetic field in a zoomed in region at a time when several plasmoids have
formed.  Figure \ref{fig:three_phases} (b) shows the energy of the electron
as a function of its position along the $x$-axis, revealing three distinct
phases to the acceleration.  The purple segment of the particle trajectory
shows the initial phase of betatron acceleration.  At early times the 
magnetic field is compressed from the colliding magnetized plasma flows,
and the magnetized electron gains a small amount of energy from adiabatic
betatron acceleration.  The electron then gains a large amount of energy in
a localized region along $x$, which can be identified as the location of a
reconnection \textit{X} point in the magnetic field profile.  Here the magnetic field
vanishes, and the electron is non-adiabatically acccelerated by the
out-of-plane electric field associated with reconnection. The in-plane
magnetic field then deflects the electron away from the \textit{X} point and into
the plasmoid, where it becomes trapped.  As new magnetic flux is added onto
the plasmoid from reconnection and the plasmoid contracts,
the electron gains an amount of energy that is comparable to that gained at
the \textit{X} point.

\begin{figure}[ht]
\begin{center}
\includegraphics[width=0.9\textwidth]{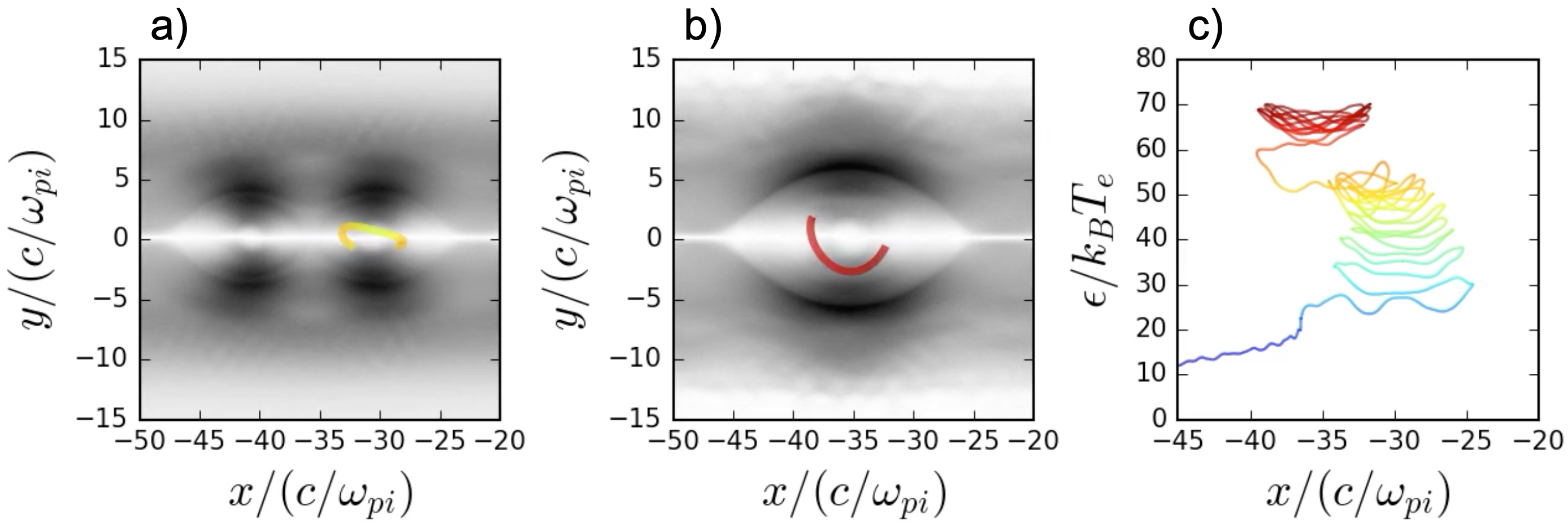}
\caption{\label{fig:plasmoid_merger} 
Magnitude of the magnetic field from the $L / d_{i} = 212$ simulation (a) before and (b) after the plasmoid
merger, with segments of the trajectory of an energetic electron, colored
by its energy according to (c).  (c) Shows the energy of the electron as
a function of its position along the $x$ axis.
}
\end{center}
\end{figure}

In addition to acceleration inside a single contracting plasmoid,
plasmoid mergers are seen to frequently occur in this system and
further energize particles.  Figure \ref{fig:plasmoid_merger} shows an
example trajectory of an electron from the $L / d_{i} = 212$ simulation, colored by its energy, before (a) and
after (b) the merging of two plasmoids.  Figure \ref{fig:plasmoid_merger}
(c) shows the electron's energy as a function of its position along the
$x$-axis.
The electron initially is trapped inside a single plasmoid and gains
energy as it contracts.  When the two plasmoids merge together, the
electron experiences a fast increase in energy as it is reflected from the
edge of the newly merged larger plasmoid.  This larger plasmoid then
continues to contract and further energizes the electron.
Acceleration associated with contracting \cite{Drake2006b} and merging
\cite{Oka2010,Le2012} plasmoids has been studied in past simulations 
and is of strong interest for explaining the efficient acceleration of
electrons and power-law energy spectra associated with reconnection
in space physics and astrophysics. In both cases the acceleration mechanism
can be interpreted as the result of Fermi Type B reflections
\cite{Fermi1949} from the magnetic field at the edge of the plasmoids.
The restriction of these simulations to two spatial dimensions makes the
plasmoids effectively uniform in the out-of-plane dimension. In
three-dimensional systems plasmoids may be unstable to the drift-kink
instability which has wave-vector components in the out-of-plane dimension.
However this instability is typically strongly suppressed over the
timescales of the
reconnection dynamics at realistic mass ratios \cite{Daughton1999a}, and
past work has found that the particle acceleration can remain
effective even in the absence of coherent plasmoids \cite{Guo2014}.
The periodic boundaries along the outflow direction in
these simulations may enhance the frequency of plasmoid mergers by
preventing plasmoids from escaping the system.  However, because the
simulations are run for less than one Alfv\'{e}n crossing time the
artificial effects of the boundaries on plasmoid evolution
and particle acceleration are limited, and nonthermal electron
acceleration and plasmoid merging are seen to persist in the
simulations of finite systems.

\begin{figure}[ht]
\begin{center}
\includegraphics[width=0.9\textwidth]{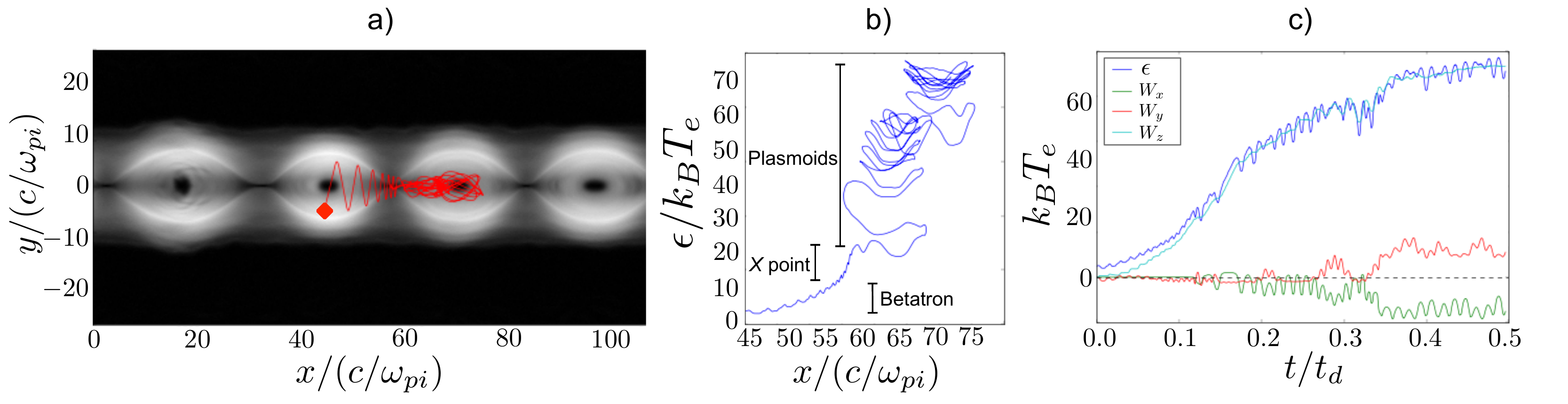}
\caption{\label{fig:electron_track} 
(a) Trajectory of an energetic electron plotted over the magnitude of the
magnetic field from the $L / d_{i} = 212$ simulation. (b) Energy of the electron as a function of position along
the $x$-axis.  (c) Total energy of the electron and work done by the
individual electric field components as a function of time.
}
\end{center}
\end{figure}

The energy gain for the processes of betatron acceleration, direct \textit{X} point
acceleration, and plasmoid contraction and merging comes from the
out-of-plane component of the electric field.  Figure
\ref{fig:electron_track} (a) shows the trajectory of another representative
energetic electron from the $L / d_{i} = 212$ simulation.  From the plot of energy as a function of position
along the $x$ axis in Figure \ref{fig:electron_track} (b), it can be seen
that this electron experiences betatron acceleration, \textit{X} point
acceleration, and acceleration from plasmoid contraction and a plasmoid
merging event.
The total energy of the particle and the work done by the three
components of the electric field are shown in Figure
\ref{fig:electron_track} (c), showing how the work done on the
electron primarily comes from the force from the out-of-plane electric field.
The energetic electron tracks we have analyzed show that the energy gain
from betatron acceleration is typically minor compared to that from the
\textit{X} point and plasmoids.  Direct acceleration by the reconnection
electric field at the \textit{X} points usually provides the first
significant gain of energy, with the remaining coming from plasmoid
contraction
and merging.  The overall energy gain from plasmoid contraction and
merging is typically comparable to or greater than that from the
\textit{X} points.  At even larger system sizes
where plasmoids evolve over longer timescales we expect the
contribution of plasmoid related particle acceleration
to become increasingly dominant, as the \textit{X} points
contribute primarily to the initial phase of acceleration.

\section{ION ACCELERATION}

\begin{figure}[ht]
\begin{center}
\includegraphics[width=0.9\textwidth]{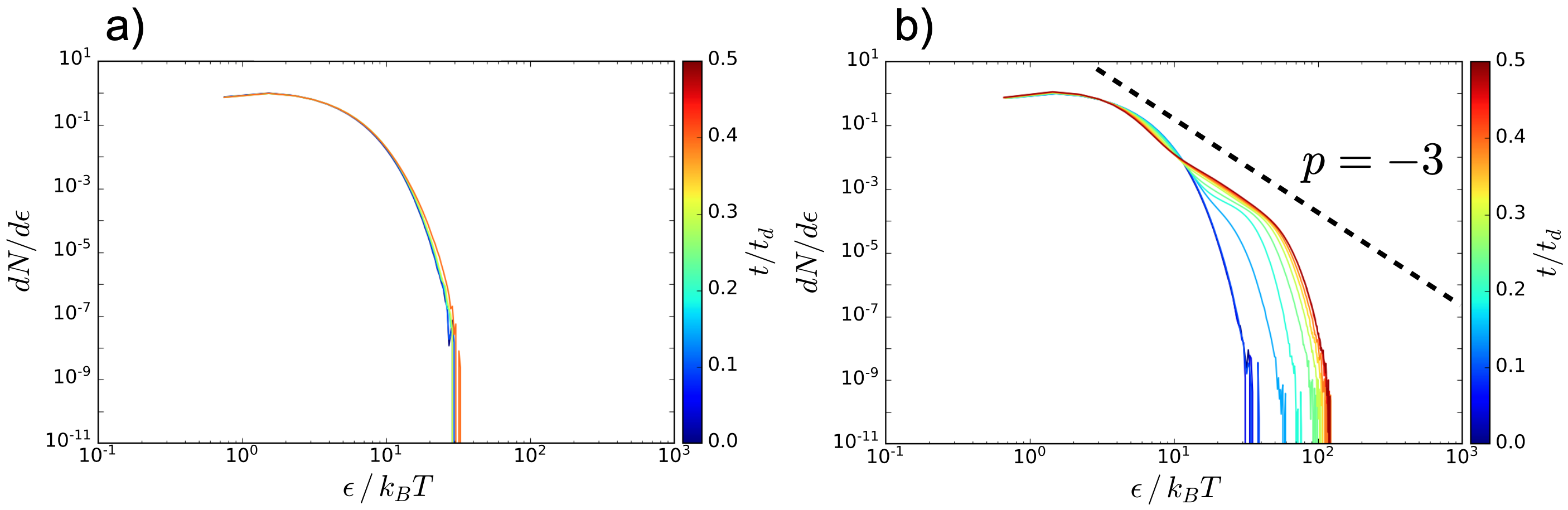}
\caption{\label{fig:ion_spectra} 
Temporal evolution of the ion energy spectra for two different system
sizes: (a) $L / d_{i} = 53$ and (b) $L / d_{i} = 212$.  The colored lines show the
evolution of the spectra from $t / t_{d} = 0$ to $t / t_{d} = 0.5$ at
equally spaced time intervals.  The nonthermal component that develops
for the larger system size resembles a power-law with an index of approximately $p=-3$ (plotted for reference).
}
\end{center}
\end{figure}

The ions exhibit a dramatic change in behavior as the system size is
increased.  Figure \ref{fig:ion_spectra} shows the temporal evolution of
the ion spectra for a system size of $L / d_{i} = 53$ in (a) and
$L / d_{i} = 212$ in (b).  For the small system size, the ions show only a
small amount of heating and remain well described by a Maxwellian.
For the large system size, the ions gain a significant amount of
energy and exhibit the clear development of a nonthermal tail in their
spectra, extending to more than 100 times the initial thermal energy.
This difference is due to the ability of the magnetic field to
magnetize the ions.  At the small system size, the amount of magnetic
flux is not sufficient to contain the ions, and the ions pass through
the interaction region before they can be energized by reconnection
related acceleration mechanisms.  At the large system size, however, ions are
magnetized and can be contained inside the reconnection region, allowing them
to be spend a significant amount of time in the reconnection layer
and become energized by reconnection.
The energization is seen to saturate at later times due to the finite
amount of plasma and magnetic flux in the system. The system is initialized
with plasma flows with a finite velocity that drive reconnection, however
because no new plasma or magnetic flux are introduced the reconnection
dynamics are strongest during the initial interaction and become weaker at
later times as the system relaxes.
Similar to what was seen for the electrons, the nonthermal component
that develops has a shape resembling a power-law, in this case
with an index of approximately $p=-3$.  The value of this index and the strong
influence of the system size on the spectrum are also 
consistent with recent satellite measurements of reconnection in
Earth's magnetotail \cite{Imada2015a}.
The energy contained above 5 (10) times the thermal energy is approximately
30 (10) percent of the total energy of the ions for the system with
$L / d_{i} = 212$.

\begin{figure}[ht]
\begin{center}
\includegraphics[width=0.9\textwidth]{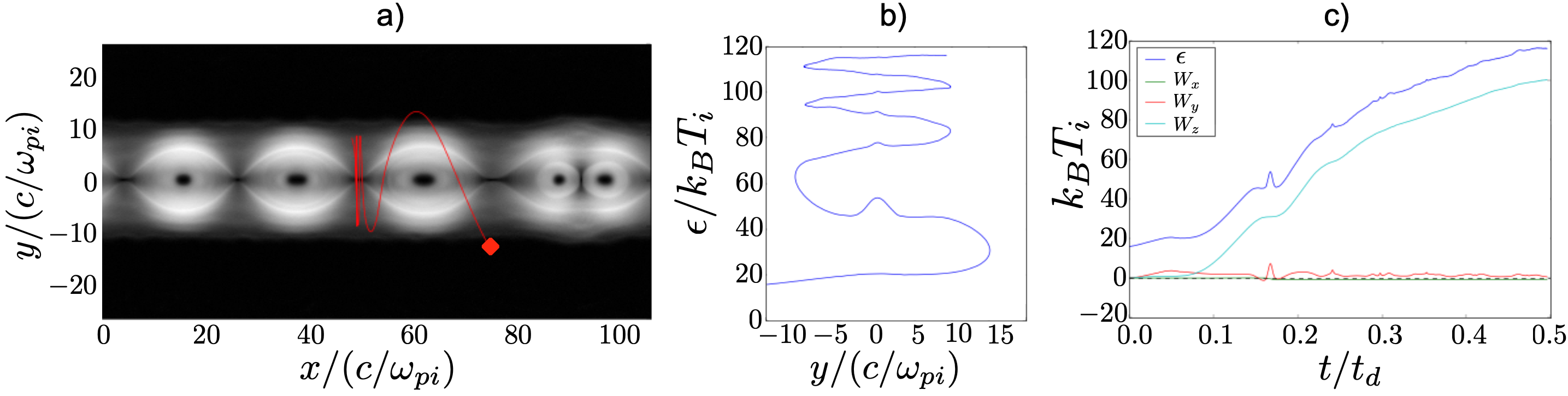}
\caption{\label{fig:ion_track} 
(a) Trajectory of an energetic ion plotted over the magnitude of the
magnetic field from the $L / d_{i} = 212$ simulation. (b) Energy of the ion as a function of position along
the $y$-axis.  (c) Total energy of the ion and work done by the individual
electric field components as a function of time.
}
\end{center}
\end{figure}

Figure \ref{fig:ion_track} shows an example evolution of an ion that
is energized by reconnection in the $L / d_{i} = 212$ simulation.  The
ion's trajectory in space is plotted over the magnitude of the magnetic
field in Figure \ref{fig:ion_track} (a), showing how it is contained in the
current sheet by the magnetic field and repeatedly passes through a
reconnection \textit{X} point.  The ion's energy as a function of its position
along the $y$ axis is shown in Figure \ref{fig:ion_track} (b), showing
how it initially gains energy from betatron acceleration during two
reflections as the magnetic field is compressed.  Once reconnection
has initiated, the ion gains energy from the reconnection electric
field as it crosses over the \textit{X} point.
Figure \ref{fig:ion_track} (c)
shows the total energy of the ion and the work done by the electric
field components, showing how it is again the out-of-plane component
of the electric field that provides the energy. For this system size,
the Larmor radii of the ions are still too large compared to the size
of the plasmoids to be energized in plasmoid related particle
acceleration mechanisms.  

\begin{figure}[ht]
\begin{center}
\includegraphics[width=0.75\textwidth]{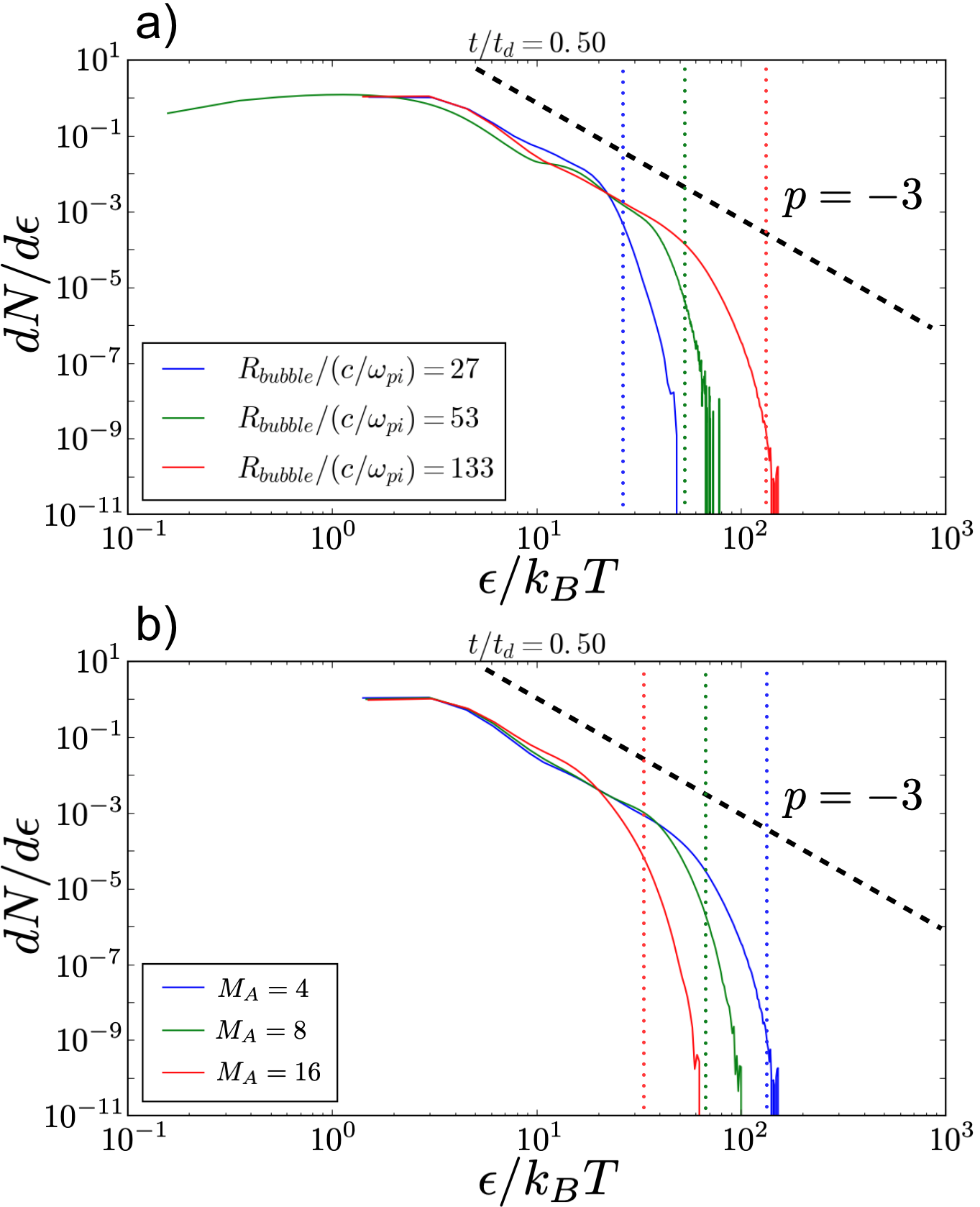}
\caption{\label{fig:spectra_comparison} 
(a) Ion spectra for finite system simulations with sizes of $R/d_{i} = 27, 53,$ and $133$. (b) Ion spectra with $R/d_{i}=133$ for three different initial magnetic field strengths, parameterized by the Alfv\'{e}nic Mach number $M_{A}$.  Blue: $M_{A} = 4$ (same strength as the spectra in (a)), green: $M_{A} = 8$,
and red:  $M_{A} = 16$.  A power-law spectrum with index $p=-3$
is plotted for reference.  Dotted lines show the maximum energy estimate
from the formula in the text for the spectrum of the corresponding color.
}
\end{center}
\end{figure}

The previous simulations were for systems that were periodic along the
direction of the reconnection outflows, which means particles can not
escape as they do in more realistic finite systems.  Figure \ref{fig:spectra_comparison} shows
the ion spectra for several simulations with finite system sizes,
demonstrating how
the ion acceleration persists in an open configuration.  Figure \ref{fig:spectra_comparison}
(a) shows a comparison
for system sizes of $R/d_{i} = 27, 53,$ and $133$, showing the
emergence of ion acceleration as the system size is increased and the
additional magnetic flux allows the ions to be contained in the reconnection
layer.  Figure
\ref{fig:spectra_comparison} (b) shows the large system size, $R/d_{i}=133$, for three
different magnetic field strengths.  The finite number of constant width
energy bins used to calculate these spectra can lead to a loss of resolution
at the low energy end, however the high energy features of interest are
well resolved. The maximum energy increases with
the initial magnetic field, consistent with the ions gaining energy
from reconnection when the magnetic flux contained is sufficient to
contain the ions.
As the magnetic field strength increases by a factor of 2, the high energy
cutoff of the distribution increases by $\approx 1.5$.  This is reduced from
the linear increase that would result from a reconnection rate and
reconnection electric field that scale linearly with the initial magnetic
field, which is expected for collisionless reconnection \cite{Pritchett2001},
due to the effects of flux pileup from the driven flows which enhances the
strength of the magnetic field in the reconnection layer.  The local
magnetic field is then stronger than the initial magnetic field in the
flows,
and this enhancement is more effective at higher Alfv\'{e}nic Mach
numbers.
Figure \ref{fig:spectra_comparison} (a) compares systems with
different sizes, while Figure \ref{fig:spectra_comparison} (b)
compares systems with different initial magnetic fields.  However,
as $V_0$ is held constant, in both cases the spectra extend to higher
energies as the amount of initial magnetic flux in the
system increases. The similarity of the shape and maximum energy of the
spectra for cases with similar amounts of initial flux shows the importance
of having a sufficient amount of flux for containing ions in the reconnection layer and
allowing them to be energized by the reconnection electric field.

The finite size of the system in the out-of-plane dimension $L_{z}$ (which
is not modeled in the 2D simulations) acts to limit the acceleration of the
particles. The maximum energy obtainable from the reconnection electric
field can be estimated by the work done by this field,
$\epsilon_{max} = q E L_{z}$, where $L_{z}$ is the length of the system
along the direction of the reconnection electric field. This was taken into
account in our previous work to estimate the maximum energy gain of an
electron in terms of the initial plasma conditions\cite{Totorica2016}, which can be
extended for ions of arbitrary $Z$ and $T_{i} / T_{e}$ as
$\epsilon_{max} / k_{B}T_{i} = (1/2) (Z T_{e} / T_{i})\left ( M_{S}^2 / M_{A} \right ) \left( L_{z} / d_{i} \right )$. In terms of the particle
gyroradius based on the inflow velocity,
$\rho_{0} = V_{0} / (q B_{0} / m c)$, this can be written as
$\epsilon_{max} = 2 m V_0^2 L_{z} / \rho_{0}$. Using the diameter of the
plasma bubbles for $L_{z}$, $T_{e} = T_{i}$, and $Z=1$ to compare with our
simulations, this estimate agrees reasonably well with the cutoffs of the
power-law spectra as shown by the dotted lines in Figure \ref{fig:spectra_comparison}. Particle
tracking of ions and electrons from the simulation with $L / d_{i} = 212$
showed that the most energetic particles travel an out-of-plane distance of
$\sim 200-300 \, d_{i}$ by the end of the simulation, comparable to the
system size.  While this maximum energy estimate was derived assuming
{\textit X} point acceleration, analyzing the energy as a function of
position along the out-of-plane direction for energetic particles in our
particle tracking shows that it is also a reasonable estimate when plasmoids
are contributing significantly to the energization.  We have previously
studied the angular dependence of energetic electron acceleration in these
experiments in detail and found that the electrons escape the system in a
fan-like profile \cite{Totorica2017}. Ions and electrons are accelerated by
the reconnection electric field in opposite directions due to the signs of
their charges, and this contrast could be a powerful experimental signature
for identifying particles that are accelerated by reconnection.

Recent reconnection experiments on NIF \cite{Fox2020} using
self-generated magnetic fields produced a long ($\sim 100 \, d_{i}$) current
sheet along the direction of the reconnection outflows, however the
smaller size ($\sim 10 \, d_{i}$) along the direction of the reconnection
electric field limited the potential for particle acceleration.  For these
conditions our maximum energy estimate gives 
$\epsilon_{max} / T_{e} \sim 3$ for electrons, which closely agrees with
the ratio of the estimated integrated reconnection electric field (based
on the measured reconnected flux and experimental timescale) to the
electron temperature predicted by simulations designed to match the experiments.
Measurements and simulations of the self-generated magnetic fields in
similar systems \cite{Gao2015} suggest that the magnetic field is limited to a region above
the target that is significantly shorter than the separation between
the centers of the plasma bubbles, indicating that this configuration is likely
not ideal for studying particle acceleration.
The use of a well-controlled external magnetic field, such as used in recent
experiments\cite{Fiksel2014a}, provides in principle a more suitable
experimental configuration where the magnetized plasma region has an
out-of-plane size that is comparable to the size of the plasma bubbles.
Assuming the alternative geometry of counter-streaming flows
\cite{Fiuza2020} and an externally imposed field of 10-30 T (producible
using a magneto-inertial fusion electrical discharge system (MIFEDS)
\cite{Fiksel2018}) significantly enhances the potential for particle
acceleration. For this case it should be possible to
have system sizes $L_{z} / d_{i} > 200$ in the direction of the
reconnection electric field, producing maximum ion energies of 
$\epsilon_{max, i} / T_{i} > 1000$ and electron energies of
$\epsilon_{max, e} / T_{e} > 100$.  Choosing configurations that
maximize the energies attainable by particles in terms of the
initial thermal energy has the advantage of extending the
power-law portion of the energy spectrum, which will help for
identifying this feature in experimental measurements and
precisely defining the spectral index.

\section{COLLISIONAL EFFECTS}

\begin{figure}[ht]
\begin{center}
\includegraphics[width=0.9\textwidth]{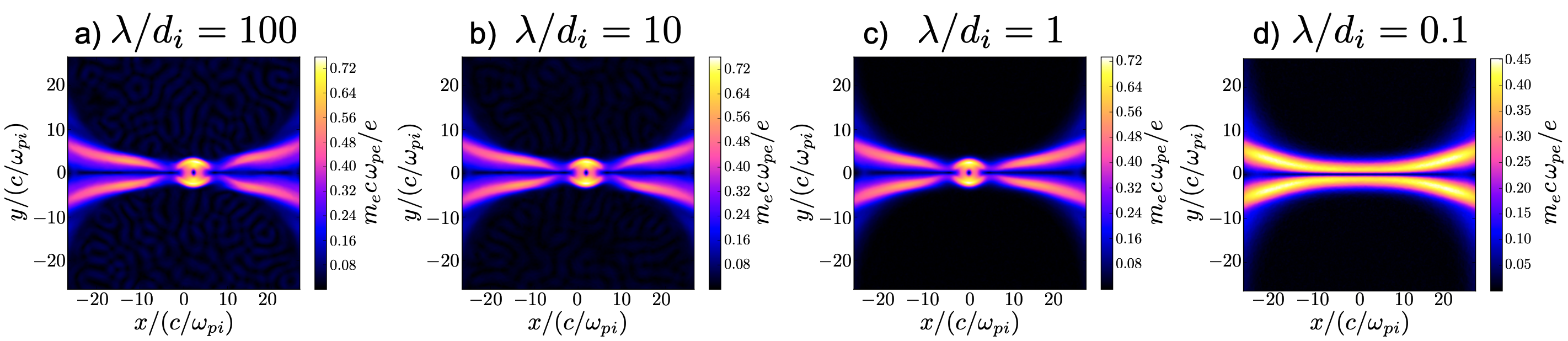}
\caption{\label{fig:plasmoid_collisions} 
Magnetic field magnitude for four simulations with the same initial
conditions ($R_{bubble} / (c/\omega_{pi}) = 27$, $M_{S} = 2$, $M_{A} = 4$) but varied levels of collisionality, parameterized by the
electron mean free path in terms of the ion skin depth $d_{i}$.  Color
scales are adjusted for each simulation to highlight the structural features of the current sheet.
(a) $\lambda / d_{i} = 100$, (b) $\lambda / d_{i} = 10$, 
(c) $\lambda / d_{i} = 1$, (d) $\lambda / d_{i} = 0.1$.
}
\end{center}
\end{figure}

Recent experimental measurements of magnetic reconnection using laser-driven
plasmas with self-generated magnetic fields did not show evidence of
plasmoid formation, and it has been argued that these results contradict the
expectations from previous simulations \cite{Rosenberg2015a,Fox2012a}.
However, these experiments were done in a regime where the collisional mean
free paths are comparable to the length scales in the system, and most
previous PIC simulations considered a collisionless plasma. It is thus
important to understand how the collisionality affects plasmoid formation
and particle acceleration in laser-driven reconnection. Collisional effects
can be included in PIC simulations using a binary Monte Carlo Coulomb
collision operator \cite{Takizuka1977}. We investigate the role of
collisional effects in these systems using PIC simulations of expanding,
magnetized plasma bubbles with varied levels of collisionality.
The Lundquist number, $S  = L V_{A} / \eta = \left ( L / d_{i} \right ) \omega_{ce} \tau_{e}$, gives the ratio of the characteristic timescale of the
magnetic field dynamics to that of resistive diffusion. Another
important parameter for considering the collisional physics in
reconnection is the electron mean free path in terms of the ion skin
depth $\lambda / d_{i}$, as $d_{i}$ is the characteristic length
scale of plasmoid formation.
This parameter is related to the Lundquist number as
$S  = \left ( L / d_{i} \right ) \left ( 2 / \beta_{e} \right )^{1/2} \left ( m_{i} / m_{e}Z \right )^{1/2} \left ( \lambda / d_{i} \right )$.  Our simulations
match the system size in terms of the ion skin depth $L / d_{i}$ and the
electron plasma beta $\beta_{e}$ to the experimental values, however
an artificially reduced mass ratio of $m_{i} / m_{e} Z = 128$ is
used.  This expression for the Lundquist number makes it evident
that when $\beta_{e}$ and $L / d_{i}$ are matched to the experimental values but an artificial mass ratio is used,
it is not possible to simultaneously match $S$
and $\lambda / d_{i}$ to the experimental values.
It remains to be determined which parameter, $S$ or
$\lambda / d_{i}$, is more important to match to accurately capture the
dynamics of the physical system with a reduced
mass ratio simulation.

We parameterize the collisionality in the system using the electron
collisional mean free path in terms of the ion skin depth.
Figure \ref{fig:plasmoid_collisions} shows the magnitude of the magnetic
field in four simulations with the same initial conditions
($R_{bubble} / (c/\omega_{pi}) = 27$, $M_{S} = 2$, $M_{A} = 4$) but varied
levels of collisionality, showing changes in the structure of the current
sheet as the collisionality changes.  By varying the electron mean free path in terms of the ion skin
depth, we see a clear suppression of plasmoid formation as this ratio drops
below one. Typical conditions for experiments with self-generated fields are
below this threshold for plasmoid formation
($\lambda / d_{i} \approx 0.5$), and thus our results suggest that the discrepancy between previous
simulations and experiments may be due to collisional effects.
Previous simulations have matched $S$ and found
plasmoid formation in a regime where experiments did not observe
plasmoids \cite{Rosenberg2015a,Lezhnin2018}.  Instead, in our simulations we match $\lambda / d_{i}$, and show that this dimensionless parameter controls the suppression of plasmoid formation as observed in recent
experiments - providing support for this approach of scaling the
reduced mass ratio simulations.
Experiments
using externally imposed fields \cite{Fiksel2014a} can produce more collisionless regimes that are above the threshold for plasmoid
formation ($\lambda / d_{i} > 10$), and are likely the ideal configuration
for studying particle acceleration and plasmoid dynamics.

\begin{figure}[ht]
\begin{center}
\includegraphics[width=0.9\textwidth]{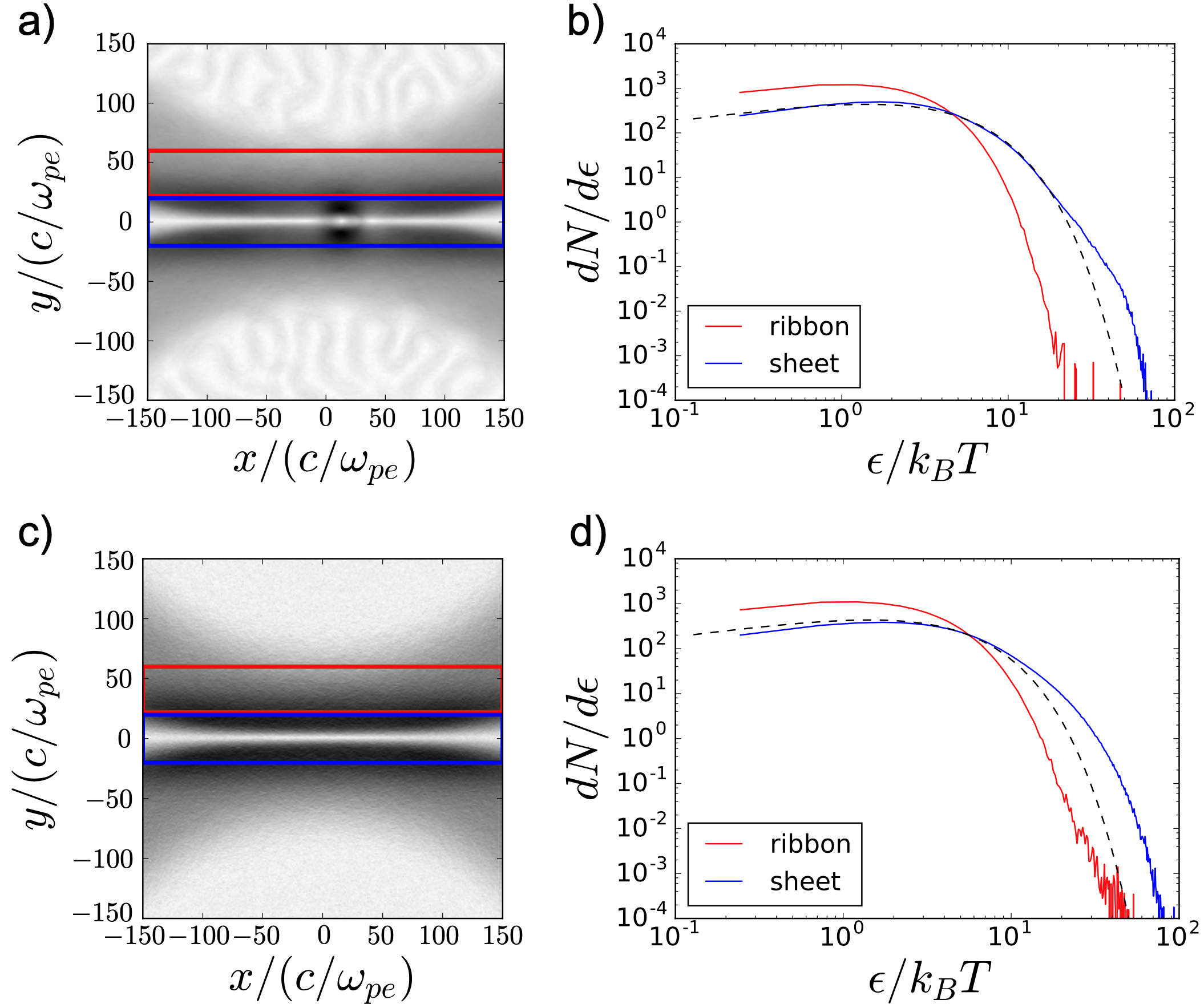}
\caption{\label{fig:collisional_spectra} 
Magnetic field magnitude for a collisionless simulation (a) and
a collisional simulation with $\lambda / d_{i} = 0.1$ (c).
(b) and (d) show the electron spectra in the regions indicated
by the boxes in (a) and (c), respectively. Dashed lines show
a Maxwell-Boltzmann distribution with the same average energy as the
spectrum inside the current sheet for the collisionless simulation. Both simulations are initialized
with $R_{bubble} / (c/\omega_{pi}) = 27$, $M_{S} = 2$, and $M_{A} = 4$.
}
\end{center}
\end{figure}

Figure \ref{fig:collisional_spectra} shows an investigation of the effects
of collisionality on the particle acceleration for simulations with
$R_{bubble} / (c/\omega_{pi}) = 27$, $M_{S} = 2$, $M_{A} = 4$.  Figures
\ref{fig:collisional_spectra} (a) and (b) show the magnetic field magnitude
and electron spectra for a collisionless simulation, and Figures
\ref{fig:collisional_spectra} (c) and (d) show the same quantities for a
collisional
simulation.  The collisional simulation has $\lambda / d_{i} = 0.1$, which
is in a regime where plasmoid formation
is suppressed and the reconnection electric field strength is below
the Dreicer field, limiting the efficiency of \textit{X} point acceleration.  The red lines in Figures \ref{fig:collisional_spectra} (b)
and (d) show the spectra just outside the reconnection
layer and the blue lines show the spectra inside the reconnection layer
(integrated over the regions indicated by the red and blue rectangles in Figures
\ref{fig:collisional_spectra} (a) and (c)). In both cases electrons are
accelerated by the reconnection electric field at \textit{X} 
points, however in the
collisional case the electrons are also heated by collisions with ions,
which only slightly changes the shape of the spectrum.
The similarity of these spectra indicates that at these smaller system sizes
the acceleration from plasmoids is not a dominant energization mechanism,
which is consistent with the findings in our previous studies
\cite{Totorica2016,Totorica2017}.  We expect the differences in the electron spectrum to become significant at larger system
sizes such as the $>100d_{i}$ 
systems studied above where significant plasmoid mediated particle
acceleration occurs in the collisionless case but would be suppressed in the collisional case.

\section{DISCUSSION}

Within the heliosphere, plasmoids can be observed directly, for example in
solar flares in the solar corona \cite{Takasao2016} and substorms in the
Earth's magnetotail \cite{Ieda1998,Chen2008}.  They are believed to have
important effects on aspects such as the reconnection rate, energy transport,
and nonthermal
particle acceleration.  Simulations for relevant conditions have shown the
acceleration of nonthermal populations of particles, but have not conclusively
determined the shape of the resulting spectra and how they depend on the plasma
conditions \cite{Oka2010,Fu2006,Hoshino2001,Drake2006b,Dahlin2014}.
In highly magnetized astrophysical objects, it is expected that the reconnection
conditions can reach the relativistic regime, where the magnetic energy per
particle exceeds the rest mass energy
($\sigma = B^{2} / 4 \pi n m c^{2}$).  Although reconnecting current
sheets cannot be directly observed for these objects, simulations for the expected
conditions show the importance of plasmoids \cite{Guo2014,Sironi2014a}.
Plasmoids are particularly important for particle acceleration, where they
provide the dominant energization mechanism contributing to the power-law
distributions of particles that are produced.
In the trans-relativistic regime of $\sigma \sim 1$, relevant for
accretion flows in black holes, nonthermal particle acceleration can
occur for both electron and ions \cite{Ball2018}.
While the conditions in these
systems are very different from those of the experiments modeled in this
study, the qualitative features of the acceleration are similar, including
contributions from both \textit{X} points and plasmoids.  The results of this
study show that laser-driven plasma experiments at large system sizes could allow
the study of reconnection in the multi-plasmoid regime in a laboratory setting
that would allow for controlled plasma conditions and tuning between different
regimes of reconnection.  Such experiments would be valuable for developing and benchmarking models of multi-plasmoid reconnection and particle acceleration that could be connected to systems in both space physics and astrophysics, due to the similarity of the physics involved and the scalability of the governing equations.

The acceleration of ions from reconnection is an important topic that has
been investigated in systems ranging from astrophysics to the laboratory
\cite{Drake2009,Drake2010a,Schoeffler2013a,Guo2016,Cazzola2016a,Li2017a,Jarvinen2018,McClements2018,Vlahos2019}.
A critical new finding in this study is the emergence of ion acceleration
at sufficiently large laser-driven system sizes.  These results are important for systems
in space physics which often feature simultaneous electron and ion
acceleration associated with reconnection events.  A particularly promising
connection can be made between these experiments and reconnection in Earth's
magnetotail. Nonthermal electrons and ions are
commonly produced in association with reconnection in the magnetotail,
however there are many unsolved problems about their
production.  How the reconnection conditions impact the acceleration
efficiency and the relative partitioning of
the dissipated magnetic energy, and why sometimes only one of the two
species experiences significant energization are aspects
that are not yet well understood \cite{Imada2011,Hoshino2018}.
Laser-driven plasma experiments capable of measuring both electron and ion acceleration by
magnetic reconnection as a function of the plasma parameters would allow a better
understanding of the similarities and differences between the acceleration of electrons
and ions. A recent study analyzed satellite observations in the magnetotail to determine
what conditions are favorable for ion acceleration and found that the spatial
system size and reconnection electric field strength are critical factors
for producing energetic ions \cite{Imada2015a}, which is directly in accordance with the
results presented above.

Additionally, the conditions in the Earth's magnetotail current sheet
are in many ways remarkably similar to those typical of laser-driven
plasma experiments, with characteristic ion temperatures of $\sim$ keV,
electron temperatures of several hundred eV, Alfv\'{e}n speed of
$\sim 1000$ km/s, and normalized system sizes on the order of hundreds of ion skin
depths. The moderate system sizes characteristic of the magnetotail
and laser-driven plasmas are unique compared to those in solar physics
and astrophysics because they are large enough to feature complex coupling
between global and microscopic scales, yet still feasible to be
modeled using fully kinetic simulations.  Studies of these systems
could be used for developing global models of reconnection that could
potentially be extrapolated to the larger system sizes characteristic
of solar physics and astrophysics.  Further exploration of the
promising synergy between reconnection in laser-driven
plasmas and the Earth's magnetotail will be the subject of future
work.

\section{CONCLUSION}

In conclusion, using fully kinetic PIC simulations, we have
modeled laser-driven plasma experiments for the conditions of future
experiments at megajoule class laser facilities such as the National
Ignition Facility.  At large system system sizes, the multi-plasmoid
regime can be reached, where many plasmoids are present simultaneously in
the current sheet and merge to larger scales.  Plasmoid related acceleration
processes enhance the efficiency of electron acceleration and the maximum
energies that can be attained.  For systems with sufficient flux to trap
ions in the reconnection layer, we find the emergence of nonthermal ion
acceleration associated with energy gain from the reconnection electric
field at the \textit{X} points.  Furthermore, using collisional PIC simulations, we find that collisional effects can suppress plasmoid formation,
and thus need to be carefully considered in the planning of experiments and numerical modeling. These results show that laser-driven plasmas
at large system sizes offer a powerful platform for
studying nonthermal electron and ion acceleration by
reconnection and the interplay between these processes in systems in 
space physics and astrophysics.

\begin{acknowledgments}

This work was supported by the U.S. Department of Energy SLAC
Contract No. DE-AC02-76SF00515 and by the U.S. DOE Early Career Research Program under FWP 100331. The authors acknowledge the OSIRIS
Consortium, consisting of UCLA and IST (Portugal) for the use of the
OSIRIS 3.0 framework and the visXD framework.  S. T. was also supported
by the NASA Jack Eddy Postdoctoral Fellowship (NNH15ZDA001N-LWS). Simulations were
run on Mira (ALCF supported under Contract No. DE-AC02-06CH1135)
through an INCITE award, on Blue Waters, and on the Bullet Cluster
at SLAC.

\end{acknowledgments}

\section*{Data Availability}

The data that support the findings of this study are
available from the corresponding author upon reasonable
request.

\bibliographystyle{apsrev4-1}
\bibliography{main}

\end{document}